\documentclass[12]{article}
\usepackage[dvips]{graphicx}

\def\beq{\begin{equation}}
\def\eeq{\end{equation}}

\def\E^b(r){{\bf e_2}}

\def\bF{{\bf F}}
\def\ba{{\bf a}}

\def\bv{{\bf v}}

\def\bx{{\bf x}}

\def\bv{{\bf v}}

\def\ba{{\bf a}}
\def\bp{{\bf p}}
\def\bv{{\bf v}}

\def\w{{\wedge}}

\begin{document}
   
\title{Principle of Local Conservation of Energy-Momentum}

\author{Garret Sobczyk   \\ 
Universidad de Las Am\'ericas - Puebla,\\ 72820 Cholula, Mexico, \\ 
email: garret.sobczyk@udlap.mx \\ \\
Tolga Yarman\\ Okan University, Akfirat, \\ Istanbul, Turkey \\ 
email: tyarman@gmail.com }
\maketitle

\begin{abstract} Starting with Einstein's theory of special relativity and the
principle that whenever a celestial body or an elementary particle,
subjected only to the fundamental forces of nature, undergoes a change in its kinetic energy then the
mass-energy equivalent of that kinetic energy must be subtracted from the rest-mass of the body or particle,
we derive explicit equations of motion for two falling bodies.
In the resulting mathematical theory we find that there are no singularities and consequently no blackholes.

\noindent{\it Subject Classification AMS 2000:} 83A05, 83D05, 83C57, 81V22.

   \end{abstract}

\section*{Introduction} 

    Special Relativity has proven itself to be an exceptionally powerful theory that has
  revolutionized human understanding of the material universe in the 20th Century \cite{eing1}, 
  \cite{E1905}, \cite{ein}.
   The purpose of the present article is to show how by imposing a {\it local conservation of energy-momentum}
 in the special theory of relativity, the theory takes on a new elegance 
  and universality.
 In \cite{TY0}, the second author considered how a single object would fall
 in the gravitational field of a celestial object infinitely more massive. 
In what follows, we derive the exact equations of motion for two bodies of arbitrary masses
falling into each other under the influence gravity. 
We find that there are no singularities, even in the case of idealized point masses. 

Section 1, defines the concept of rest-mass utilized in our theory. 
Whereas Einstein, by his {\it equivalence principle}, 
considers ``inertial mass'' and ``rest-mass'' to be equivalent, we believe that there
 is a clear asymmetry between an
 accelerating elevator and a gravitational field.
 An observer must get {\it accelerated} to be able
 to catch up with an accelerating elevator, whereas he has to get {\it decelerated} in order to be able
 to land on the celestial body. In our theory, the first process yields a {\it mass 
 increase}, whereas the second one leads to
 a {\it mass decrease} \cite{typl}.  It follows that 
the idea that the rest-mass of an object is a fundamental constant of nature, must be replaced by
the concept of the {\it instantaneous rest-mass} of an object in a 
non-homogeneous field, as was first done in \cite{TY0}.   

Section 2, defines the concept of {\it binding energy} of a two body system to account for the
work done by any one or all of the four fundamental forces of nature. 
 We find explicit formulas both for the masses and also for the velocities of the two masses
in terms of the total binding energy.
All our calculations are based upon the simple principle that each body, as it moves under the
forces of nature, must subtract the mass-equivalent for any change in its kinetic energy. 
We express the ideas of special relativity in the 
 framework of the {\it Minkowski spacetime algebra} (STA) developed by D. Hestenes \cite{H74}. In STA,
 each relative frame of an observer is defined by a unique, future pointing, Minkowski 
timelike unit vector tangent to 
the timelike curve called the {\it history} of that observer. 

Section 3, considers the binding energy due to Newton's gravitational force between two 
bodies and derives a Riccati-like differential equation of motion. 
We find closed form solutions for the case of a celestial body and for the case when two bodies have the same mass.
In the general case when an exact solution is not possible, we use a numerical solution.  

Section 4, discusses the results of previous sections and concludes
 that black holes with a well defined Schwarzschild radius cannot exist. 

  \section{The concept of rest-mass}
  
  We begin by defining the {\it rest-mass} $m_\infty$ of a body to be the mass of the body  
when it is isolated from all other bodies and forces in the Universe, as 
measured by an observer traveling at relative rest with respect to that body. 
The great advantage of the STA of Hestenes, for the most part still unappreciated by the physics community,
is that each such inertial frame is uniquely characterized
by a constant Minkowski time-like unit vector $u$. 
 See \cite{H74} and \cite{H03} for 
   details of the spactime algebra formulation of special relativity which we use throughout this paper.
   
      Let $p_\infty$ be the {\it Minkowski energy-momentum vector} of the rest-mass $m_\infty$. 
      Since we have assumed that $m_\infty$
  is at rest in the frame defined by $u$, it follows that $p_\infty=m_\infty c^2 u$.   
  Now let
  $v=\frac{dx}{d\tau}$ be the Minkowski timelike unit vector of an observer with
the timelike history $x=x(\tau)$, where
 $\tau$ is the natural parameter of {\it proper time} (arc length). The unit vector $v=v(\tau)$ uniquely defines the
 {\it instantaneous frame} of the observer at the proper time $\tau$. 

As measured from the rest-frame $u$ to the instantaneous relative frame $v$, we have
    \beq  p_\infty v = m_\infty c^2 uv = m_\infty c^2 (u\cdot v+  u\w v) 
    =\gamma_v m_\infty c^2(1+ \frac{\bv}{c}),\label{instantv}  \eeq
  where $\gamma_v = u\cdot v=\frac{1}{\sqrt{1-\frac{\bv^2}{c^2}}}$ 
  and $\frac{\bv}{c} = \frac{u\w v}{u\cdot v}$. We say that
  $E_v=p\cdot v=\gamma_v m_\infty c^2$ is the instantaneous {\it relative energy},
 $\bp_v=\gamma_v m_\infty c^2 \frac{\bv}{c}$
  is the instantaneous {\it relative momentum}, and 
  $\bv$ is the instantaneous {\it relative velocity} of $m_\infty$ in the instantaneous frame $v$ 
as measured by $u$. This convention is opposite by a sign 
  to the convention used by Hestenes in his 1974 paper. We use the same convention here as used by Sobczyk
  in \cite{S1}. There are many different languages and offshoots of languages that have been used to formulate
  the ideas of special relativity. For a discussion of these and related issues, see 
  \cite{BS}, \cite{S3}. A unified language for mathematics and physics has been proposed in
  \cite{H/S}. 
  
  Equation (\ref{instantv}) shows that with respect to
 the relative frame $v$, the mass $m_\infty$ has the increased relative
  energy $E_v=\gamma_v m_\infty c^2$. This means that if we want to boost the mass $m_\infty$ 
  from the rest-frame $u$ into the instantaneous frame $v$, we must expend the energy
  $ \triangle E_1=(\gamma_v-1)m_\infty c^2$ to get the job done. Expanding the right-hand side of this
  last equation in a Taylor series in $|\bv|$, we find that 
  \beq \Delta E_1=\frac{m_\infty }{2}\bv^2+ \frac{3 m_\infty}{8 c^2}\bv^4+\frac{5m_\infty}{16c^4}\bv^6+\cdots .
                        \label{tayexpand1} \eeq
 For velocities $|\bv|<< c$, we see that the energy expended to boost the mass $m_\infty$ into the instantaneous
 frame $v$ moving with velocity $\bv$ with respect to the rest-frame $u$ is $\Delta E_1\widetilde 
 =\frac{m_\infty}{2}\bv^2$,
 which is the classical Newtonian expression for {\it kinetic energy} of the mass $m_\infty$ moving with
 velocity $|\bv|$.                       
  
  If, instead,
  we pay for the work done by deducting the required energy-equivalent from the mass $m_\infty$, to get
  the residual rest-mass $m=\frac{m_\infty}{\gamma_v}$, then
  the terminal energy-momentum vector of the mass $m_\infty$ when it has reached the velocity $\bv$ is
     \beq p=mc^2 v=\frac{m_\infty }{\gamma_v}c^2 v =\frac{p_\infty}{\gamma_v}uv
      =e^{-\frac{\phi \hat \bv}{2}}\frac{1}{\gamma_v}p_\infty e^{\frac{\phi \hat \bv}{2}}.
            \label{subenergy}   \eeq
 In this equation, $\hat \bv$ is a unit relative vector in the direction of the velocity $\bv$, and
 $c \tanh(\phi)=|\bv|$ is the magnitude of the velocity as measured in the rest-frame $u$.

 Equation (\ref{subenergy}) has some easy but important consequences. We first note that
 $m=\frac{m_\infty}{\gamma_v}=0$ when $|\bv|\rightarrow c$. This means that the energy content of each material
body is exactly the energy which would be required to accelerate the body to the speed of light $c$. Assuming
that we have a one hundred percent efficient photon drive, the body would reach the speed of light at precisely
the moment when its last bit of mass-equivalent is expelled as a photon. A second interesting observation is
 that when we expand $(m_\infty-m)c^2=m_\infty(1-\frac{1}{\gamma_v})c^2$ in a Taylor series in $|\bv|$ around
$|\bv|=0$, we obtain
   \beq \Delta E_2=( m_\infty-m)c^2=\frac{m_\infty }{2}\bv^2+ \frac{m_\infty}{8 c^2}\bv^4+
   \frac{m_\infty}{16c^4}\bv^6+\cdots =\frac{\Delta E_1}{\gamma_v}.
                        \label{tayexpand2} \eeq
 Whereas the expressions $\Delta E_1\widetilde = \Delta E_2$ for $|\bv|<<c$, the expression
 for $\Delta E_2$ is much closer to the classical kinetic energy over a much larger range of velocities
 $|\bv|<c$, and differs only by a factor of $2$ when $|\bv|=c$.                       
  
    The basic premise upon which our theory is built is 
    that when any particle evolves on its {\it timelike curve} $x(\tau)$,
  subjected only to the elementary forces of nature and satisfying the initial condition that
  $p(0)=m_\infty c^2 u$, then its energy-momentum vector has the form $p(\tau)=m(\tau)c^2 v(\tau)$
  for $m(\tau)=\frac{m_\infty}{\gamma_v}$, and
  satisfies the {\it conservation law} 
     \beq p(\tau)\cdot u = m_\infty c^2 = constant  \label{nature}  \eeq
for all values $\tau \ge 0$. This law is  
a direct consequence of the local conservation of energy requirement (\ref{subenergy}).
We say that 
  \beq m(\tau)=\frac{\sqrt{ p^2}}{c^2} =\frac{p(\tau) \cdot v(\tau)}{c^2}=\frac{m_\infty}{\gamma_{v}}  \label{instantmass} \eeq
  is the {\it instantaneous rest-mass}
of $m_\infty$ in the instantaneous frame $v(\tau)$. 

       At the atomic level, our insistance upon the strict
  {\it local conservation of the energy-momentum} of each particle (\ref{nature}), 
   means that whenever an elementary particle undergoes a change in
  its kinetic energy, it must pay for it with a corresponding change in its instantaneous rest-mass (\ref{instantmass}).
  Thus, we do not accept that the rest-mass $m_\infty$ of an isolated particle is an invariant when
  that particle undergoes interactions. 
     We consider that the {\it field} of
   an elementary particle carries only {\it information} about the {\it location} of that elementary particle, but does not
   magically transfer energy across spacetime to affect other elementary particles. Each elementary particle
    pays for any change in its kinetic energy as it navigates in space, guided by
   the information supplied by the four elementary forces of Nature. Consequently, an elementary
   particle annihilates if and only if it reaches the speed of light.
   
   A beautiful discussion and derivation of the basic relationships of relativistic particle dynamics is 
   given in \cite{H03}    
and \cite{H99}, so we need not rederive them here. We will need, however,
a number of special formulas regarding the evolution of a particle
 whose the energy-momentum vector is given by 
$p(\tau)=m(\tau)c^2 v(\tau)$ and satisfies
(\ref{nature}), as given above. The {\it Minkowski force} on such a particle as it moves along its
timelike curve $x(\tau)$, is given by $f(\tau)= \frac{d p(\tau)}{d \tau}$.
 It is very easy to calculate the {\it relative force} $\bF(\tau)=\frac{1}{c^2}u f(\tau)$ as 
 measured in the rest frame $u$. We find that
   \beq   \bF(\tau)=\frac{1}{c^2}u f(\tau) =\frac{1}{c^2} \frac{d u p(\tau)}{d \tau}=
   \frac{dt}{d\tau}\frac{d}{dt}(m_\infty +m_\infty  \bv)
       = \gamma_v m_\infty  \ba,  \label{force} \eeq
  where $\ba = \frac{d\bv}{dt}$ is the {\it relative acceleration} experienced by the particle as measured in the
  rest frame $u$. Formula (\ref{force}) is immediately recognized as the relativistic form of {\it Newton's Second Law}.
  This form of Newton's Second Law applies to particles subjected only to elementary forces. 
  Noting that $\frac{1}{\gamma_v^2}=1-\frac{\bv^2}{c^2}$, so that
    $ \frac{d}{dt}(\gamma_v^{-2})=-2 \frac{\bv\cdot \ba}{c^2}$, it is easy to calculate the useful formulas
    \beq \frac{d\gamma_v }{dt}=\gamma_v^3 \frac{\bv\cdot \ba}{c^2} \label{dgamma} \eeq
  and, with the help of (\ref{force}),
    \beq \frac{d m(\tau) }{d\tau}=\gamma_v \frac{d m(\tau) }{dt}=-\gamma_v^2 m_\infty  \frac{\bv\cdot \ba}{c^2}
               =- \frac{\gamma_v}{c^2}\bF\cdot \bv , \label{dmass} \eeq
  or 
     \beq \frac{d m(\tau) }{d t}=   - \frac{1}{c^2}\bF\cdot \bv . \label{dmasst} \eeq          
    
       It is well-known that the {\it total energy-momentum vector} of an isolated $n$-particle system
 is a constant of motion in every inertial system \cite[p. 634]{H99}. Assuming that the only
 interactions between the particles are the elementary forces, so that (\ref{nature}) applies, 
 it follows that the energy-momentum
 vector of each particle has the form $p_i(t)=m_i(t)c^2 v_i(t)$, and $p_i(0)=m_i^\infty c^2 u$, where  
  $t$ is the parameter of relative time in the rest-frame $u$. Assuming further that there are no collisions, 
  this conservation law takes the form  
   \beq P(t)=\sum_{i=1}^n p_i(t)=P_0 = \sum_{i=1}^n p_i(0) \label{hconservation} \eeq
for all $t\ge 0$. Dotting and wedging each side of this equation on the left by $u$, gives the
equivalent statements that
    \[u \cdot P(t) = \sum_{i=1}^n m_i^\infty c^2 =u \cdot P_0, \]
meaning that the {\it total energy} of the isolated system is constant,    
 and that the {\it total linear momentum}  
    \[ u\wedge P(t)= \sum_{i=1}^n m_i^\infty c^2 \bv_i(t) =u \w P_0 =0 \]
 of the isolated system is $0$ for all values of $t\ge 0$.   
      
   \section{Change of mass due to binding energy}
     
   Let us consider an isolated system of two objects $m_i(r)$, with the respective energy-momentum vectors
   $p_i(r)=m_i(r) c^2 v_i(r)$, for $i=1,2$, when they are a distance $r$ from each other as measured
   in the rest-frame $u$. This means that the objects can only interact with each other, 
   and that they begin at rest in the rest-frame $u$ when $r=\infty$. Thus,
    $\lim_{r \to \infty}p_i(r)=m_i^\infty c^2 u$ for $i=1,2$.

     The conservation law (\ref{nature}) and
   the conservation law of total energy-momentum (\ref{hconservation}) applied to our two particle system gives
       \beq P^\infty =p_1^\infty+p_2^\infty =p_1(r)+p_2(r)=P(r) \label{energyM} \eeq
  for all values of $r\ge 0$. Equivalently, 
   \[ u \cdot P^\infty = (m_1^\infty +m_2^\infty)c^2 =u \cdot P(r),       \]
 which is the conservation of the total energy of the system for all $r\ge 0$,
   and
    \beq  0= \frac{u \w P^\infty}{c^2} =\frac{u \w P(r)}{c^2} = m_1^\infty \bv_1(r)+m_2^\infty \bv_2(r), 
    \label{linearM} \eeq
which is the conservation of the total linear momentum of the system for all $r\ge 0$.
    
     The quantities 
           \beq E_i^b(r) = p_i(r)\cdot (u-v_i(r))= m_i^\infty c^2(1-\frac{1}{\gamma_i}),
            \label{bindingcondition}  \eeq
   which are seen in (\ref{tayexpand2}) to be closely related to the classical kinetic energy,
   are called (by the first author) {\it Tolga's binding energies}
  of the respective bodies $m_i(r)$ when they are brought {\it quasi-statically} (very slowly)
   to a distance $r$ from each other in the rest-frame $u$.   
   The {\it total binding energy} $E^b(r)=E_1^b (r)+E_2^b (r)$, is the work done by the gravitational attraction
   between the two bodies. 
    With the help
   of formula (\ref{dmasst}), we can easily calculate
   \beq \frac{dE^b}{dt}=-c^2(\frac{dm_1(\tau_1)}{d t} +\frac{dm_2(\tau_2)}{d t})=\bF_1\cdot \bv_1+
               \bF_2 \cdot \bv_2=\frac{dE^b}{dr}\frac{dr}{dt} . \label{dbinding} \eeq
    Whereas we are only
  interested here in the binding energies of the two bodies due to the force of gravity, all our considerations
  can be applied more broadly \cite{typl}.
  
                Let us directly calculate the change of the rest-masses $m_1^\infty$ and $m_2^\infty$ as
      the two masses move under the force of gravity. 
      Very simply, the instantaneous rest-masses $m_i(E_i^b)$ are specified by
         \beq m_i(E_i^b) = m_i^\infty - \frac{E_i^b}{c^2},   \label{quasi-staticmass} \eeq
 where $E_i^b$ is the instantaneous binding energy of $m_i^\infty$, 
         as follows directly from the binding condition (\ref{bindingcondition}). The
       total binding energy between the
       instantaneous rest-masses $m_1(E_1^b)$ and $m_2(E_2^b)$ is given by
         $E^b=E_1^b+E_2^b$ .
          For our considerations below, we
       will assume that $m_2^\infty = s m_1^\infty$ for a constant value of 
       $s\ge 1$, so that $m_2^\infty \ge m_1^\infty$.
         
       Because of the total binding energy $E^b$ expended by the forces acting between them, as measured in
       the rest-frame $u$, the bodies will have gained the respective velocities $\bv_1(E^b)$ and $\bv_2(E^b)$, 
        fueled by the respective losses to their rest-masses $m_1^\infty$ and $m_2^\infty$. Precisely,
       we can say that
         \beq m_i^\infty - f_i \frac{E^b}{c^2}=\frac{m_i^\infty}{\gamma_i} \label{basicequation} \eeq
       where $f_i$ is the fraction of the total binding energy $E^b$ given up by $m_i^\infty$ for $i=1,2$, respectively.  
       This means that $f_1+f_2=1$, and, by the conservation of linear momentum (\ref{linearM}), we also know that
    $(m_1^\infty)^2 \bv_1^2=(m_2^\infty)^2 \bv_2^2$ or $\bv_2^2=\frac{1}{s^2}\bv_1^2$. Using this information,
    leads to the system of equations
      \beq m_1^\infty \Big(1-\sqrt{1-\frac{\bv_1^2}{c^2}} \Big) - f_1 \frac{E^b}{c^2} = 0 \ \ {\rm and} \ \
          m_1^\infty \Big(s-\sqrt{s^2-\frac{\bv_1^2}{ c^2}} \Big) - (1-f_1) \frac{E^b}{c^2} = 0.   
                                                         \label{systemequations} \eeq 
       
      Solving the sytem of equations (\ref{systemequations}) for $f_1$ and $\bv_1^2$ in terms of the
      binding energy $E^b$, we find that
      \[   f_1(E^b)=1-  \frac{m_1^\infty s c^2}{E^b}+ 
      \frac{2  {m_1^\infty}^2 s
   (s+1) c^4-2  {E^b}  {m_1^\infty} (s+1)
   c^2+ (E^b)^2}{2 E^b ({E^b}-c^2  {m_1^\infty}
   (s+1))}
    \]
   and ${\bv_1^2(E^b)}$
   \beq 
   =-\frac{ {E^b} \left( {E^b}-2 c^2
    {m_1^\infty}\right) \left(4  {m_1^\infty}^2 s (s+1)
   c^4-2  {E^b}  {m_1^\infty} (2 s+1)
   c^2+ (E^b)^2\right)}{4 c^2  {m_1^\infty}^2
   \left( {E^b}-c^2  {m_1^\infty}
   (s+1)\right)^2}. \label{v12} \eeq
   In the interesting special case when $m_2^\infty = s m_1^\infty$ and $s \to \infty$, we find
   that the velocity 
      \beq \bv_1^2 \to \frac{E^b(2c^2 m_1^\infty - E^b)}{c^2(m_1^\infty)^2} . \label{v11infinity}  \eeq
We will use this result later.
         
   Similarly, we can now obtain the instantaneous rest-masses 
   \[m_1(E^b)=m_1^\infty (1-f_1 \frac{E^b}{m_1^\infty c^2})\]
   or
      \beq m_1(E^b)= 
    {m_1^\infty}(1+s) -\frac{E^b}{c^2}-\frac{2  {m_1^\infty}^2 s (s+1) c^4-2  {m_1^\infty}
   (s+1) E^b c^2+ (E^b)^2}{{2 c^2}(E^b-c^2  {m_1^\infty}
   (s+1))}              \label{m1} \eeq
    and $m_2(E^b)=sm_1^\infty (1-(1-f_1) \frac{E^b}{sm_1^\infty c^2})$ or 
    \beq m_2(E^b)=\frac{2 {m_1^\infty}^2 s (s+1) c^4-2
   {m_1^\infty} (s+1) E^b c^2+(E^b)^2}{{2 c^2}(E^b-c^2 {m_1^\infty}
   (s+1))}.   \label{m2} \eeq
   
      We now calculate for what {\it critical value} $E_c^b$ of the 
  binding energy $E^b$ the smaller mass $m_1(E_c^b)=0$. We find that
      \[ E_c^b=c^2 {m_1^\infty} \left(s+1-\sqrt{s^2-1}\right). \]
   For this value of the binding energy $E^b$, we find that
    \[ m_2(E_c^b)={m_1^\infty}\sqrt{s^2-1},\ \  \bv_1^2(E_c^b) = c^2, \ \ {\rm and } \ \
    \bv_2^2(E^b)=\frac{\bv_1^2(E^b)}{s^2}.  \]
     We also find that $f_1(E_c^b)=\frac{1}{1+s-\sqrt{s^2-1}}$.
\begin{figure}
\begin{center}
\includegraphics[scale=1.50]{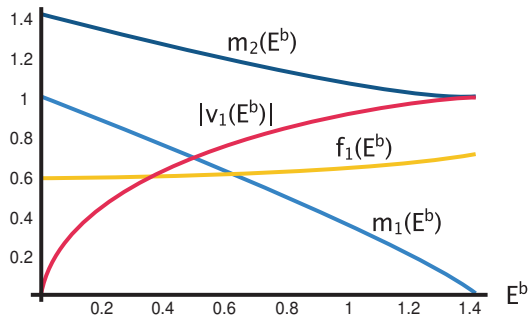}
\caption{The masses $m_1(E^b)$ and $m_2(E^b)$, the velocity $|\bv_1(E^b)|$ and $f_1(E^b)$ are plotted as functions of
the binding energy $E^b$. Initially, $m_1(0)=1$, and $m_2(0)=\sqrt{2}$.}
\end{center}
\end{figure}                                                          
           
 It is interesting to graph the instantaneous rest-masses $m_i(E^b)$ for
     $i=1,2$, the velocity $|\bv_1(E^b)|$ and the fraction $f_1(E^b)$ of 
     the binding energy being consumed by the first mass, in terms of the total binding energy $E^b$ being
     expended. In Figure 1, the velocity of light $c=1$, the mass $m_1^\infty = 1$, $m_2^\infty = \sqrt{2}$, and
     the binding energy $E^b$ satisfies the constraints $0 \le E^b \le \sqrt{2}$. At the 
     critical value $E^b=\sqrt 2$ the mass $m_1^\infty$ has entirely consumed itself. Note that 
     up to now, we have made no assumption
     regarding the nature of the force or forces which produce this binding energy. In the next section,
     we will assume that the binding energy is due to an inverse square law attractive force such as that
     due to Newton's law of gravitational attraction.   
    
      \section{Binding energy due to Newton's gravitational force} 
      
      Current knowledge tells us that there are four fundamental forces in Nature acting 
  between the two objects.   We will consider here only the force due to gravitational
  attraction between the two objects with histories $x_i(r)$ and the respective Minkowski energy-momentum
  vectors $p_i(r)=m_i(r)c^2 v_i(r)$, where the instantaneous rest
  masses are given by $m_i(r)=\frac{m_i^\infty}{\gamma_i(r)}$, and where
  $r$ is the distance between their centers as measured in the rest frame $u$. 
  
  Thus, the two bodies $m_1(r)$ and
  $m_2(r)$, in their respective instantaneous frames $v_1(r)$ and $v_2(r)$
   at a distance of $r$, will experience a mutually attractive force 
    \beq F = \frac{G m_1(r)m_2(r)}{r^2}=\frac{G}{c^4}\sqrt{\frac{p_1^2p_2^2}{(x_1-x_2)^4}},   \label{Tolga}  \eeq 
  where $G=6.67\times 10^{-11}N \, \frac{m^2}{kg^2}$ is
  Newton's constant. It is worth recalling that 
    \[r=|\bx_1-\bx_2|=\sqrt{[(x_1-x_2)\w u]^2}=\sqrt{-(x_1-x_2)^2},  \]
  since $(x_1-x_2)\cdot u = ct-ct=0$ for the simultaneous events $x_1(r)$ and $x_2(r)$ at the time $t$
  as measured in the rest frame $u$. The fact that we can express (\ref{Tolga}) entirely
  in terms of the energy-momentum vectors $p_i$ and the histories $x_i$ implies
  that Newton's Law is {\it Lorentz invariant}. An explanation of how the
  $\frac{1}{r^2}$ dependency of Newton's Law becomes a requirement of special
  relativity can be found in \cite{typl}.

 \begin{figure}
\begin{center}
 \includegraphics[scale=1.50]{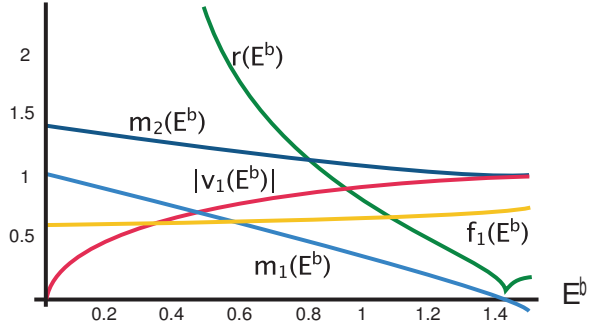} 
\caption{This figure is the same as figure 1, except that the numerical inverse solution
$r(E^b)$ for the distance $r$ between the two bodies, acted upon by the force of gravity, is shown as a function of the
binding energy $E^b$.
 Note that the value of $r\to 0$ at exactly the moment the binding energy $E^b=\sqrt{2}$,
 and that $\lim_{E^b \to 0} r(E^b)=\infty$. }
\end{center}
\end{figure} 

   In the case that the binding energy between the two bodies is totally due to Newton's
  gravitational attraction (\ref{Tolga}), we can write down the differential equation for
  the total binding energy $E^b(r)$ as a function of the distance $r$ between the two bodies
  as measured in the rest-frame $u$. We get
    \beq \frac{dE^b}{dr} =- \frac{G m_1(E^b(r))m_2(E^b(r))}{r^2}   \label{deqnewton}  \eeq         
where $m_1(E^b(r))$ and $m_2(E^b(r))$ are given in (\ref{m1}) and (\ref{m2}), respectively. 
Making these substitutions, we arrive at the rather complicated Riccati-like differential equation
  \[4 G  {(m_1^\infty)}^4 s (s+1)^2 (2 s+1) c^8-4 G
    {(m_1^\infty)}^3 (s+1) \left(5 s^2+6 s+1\right)
    {E^b}(r) c^6   \] 
    \[ +2 G  {(m_1^\infty)}^2 \left(11 s^2+18
   s+7\right)  (E^b(r))^2 c^4\] 
   \[ -12 G  {(m_1^\infty)}
   (s+1)  (E^b(r))^3 c^2+3 G
    (E^b(r))^4  \] 
   \[ +\left(-4  {(m_1^\infty)}^2 r^2 (s+1)^2
   c^8+8  {(m_1^\infty)} r^2 (s+1)  {E^b}(r) c^6 
    -4 r^2
    (E^b(r))^2 c^4\right)  {E^b}'(r)\] 
    \[=0. \]    
We shall consider the solutions of various special cases of this differential equation.

  We first consider a numerical solution in the case that $m_1^\infty=1$, $m_2=s=\sqrt{2}$,
  and the constants $G=c=1$. For this case, the graph of the solution is given in Figure 2. 
  Note that we are actually plotting the {\it inverse function} $r(E^b)$ of the solution. This
  is permissible because, as can be seen in the figure, $r(E^b)$ is a {\it strictly decreasing} function
  in the physical range of interest for $0 < E^b \le \sqrt{2}$. Note also that $r(\sqrt{2})=0$,
  although the accuracy of the numerical solution does not clearly show this.

   In the case that the body $m_2^\infty$ is so massive that $m_2(r)=m_2^\infty$ for all values of $r\ge 0$,
  the differential equation (\ref{deqnewton}) becomes 
     \beq  \frac{dE^b}{dr} = -m_2^\infty \frac{G m_1(r)}{r^2}, \label{celestial}  \eeq
   which, together with the boundary condition that $E^b(\infty)=0$, gives the particularly surprising solution
     \[ E^b(r)=E_1^b(r)=c^2(1-e^{-\frac{Gm_2^\infty}{c^2r}})m_1^\infty, \]
   or solving (\ref{quasi-staticmass}) for $m_1(r)$,
     \[ m_1(r)= e^{-\frac{Gm_2^\infty}{c^2r}}m_1^\infty .  \]
   Using (\ref{v11infinity}) and the expression for $E^b(r)$ above, we find the velocity
    \[ |\bv_1(r)|= c\Big(1-e^{\frac{-Gm_2^\infty}{c^2r}}\Big). \]
  See Figure 3. The differential equation (\ref{celestial}) and its solution, were first derived in \cite{typl},
  and a discussion of how it is related to the total energy found by Einstein can be found therein. 
               
    \begin{figure}   
  \begin{center}
\includegraphics[scale=1.50]{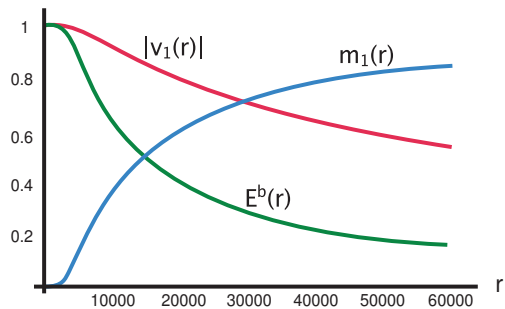}
\caption{The mass $m_1(r)$, the binding energy $E^b(r)$, and the velocity $|\bv_1|$ are shown for $0 \le r \le 60000$.
This is the case of binding to a celestial body. To make this figure, we have
assumed that $m_2^\infty=10000 m_1^\infty$ where $m_1^\infty =1$. }
\end{center}
\end{figure}     
 
    Another interesting two body case is when the masses $m_1^\infty = m_2^\infty$. 
 In this case
     the differential equation for the binding energy becomes
     \beq \frac{dE^b(r)}{dr}=2 \frac{dE_1^b}{dr}=-\frac{Gm_1^2(r)}{r^2}=
     -\frac{G(m_1^\infty -\frac{E_1^b(r)}{c^2})^2}{r^2} ,\label{equalbodies} \eeq
     which has the simple solution
       \[  E^b(r)=\frac{2c^2 G (m_1^\infty)^2}{Gm_1^\infty +2c^2r}.   \]
     We also easily find
       \[  m_1(r)=m_1^\infty-\frac{E_1^b(r)}{c^2}=\frac{2c^2m_1^\infty r}{Gm_1^\infty+2c^2 r}, \]
    and using (\ref{v12}), the velocity 
      \[  |\bv_1(r)| =c \frac{\sqrt{Gm_1^\infty (Gm_1^\infty+4 c^2 r)}}{G m_1^\infty+2c^2 r}  . \]   
   See Figure 4. The terminal velocities of the equal bodies $m_1(r)$ and
   $m_2(r)$, when they self-annihilate, are equal to the speed of light $c$.   
   
    Figures 3 and 4 strongly suggest that black holes do not exist. Whenever a light mass
     approaches an extremely dense object, depending upon initial conditions, it will necessarily
     self-annihilate or coalesce. There cannot be 
     any critical mass which would define the {\it  Schwarzschild radius} of a black hole. 
     
               \begin{figure}   
  \begin{center}
\includegraphics[scale=1.50]{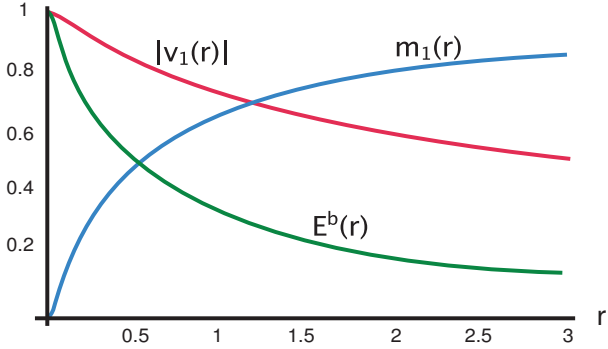}
\caption{The mass $m_1(r)=m_2(r)$, the binding energy $E_1^b(r)$, and the velocity
  $|\bv_1(r)|$ are shown for $0\le r \le 3$. }
\end{center}
\end{figure}
      
\section{Discussion}   
 
  A major problem of general relativity is that it does not easily lend itself to
quantization, although Einstien, himself, apparently did not believe in quantum mechanics \cite{EPR}.
 We have seen that, theoretically, when a mass falls from infinity into a larger mass it will self-annihilate
at $r=0$. However, quantum mechanics implies that the object's dimensions effectively  
become that of space itself at $r=0$. The restrictions of quantum mechanics imply, therefore,
that $r$ can never reach the value $r=0$ for macroscopic objects. In the case of
elementary particles, where additional forces other than gravity are known to be at work,
 self-annihiliation {\it does} occur. We have already seen that in our approach singularities,
 even those arising from the inverse square dependency of Newton's Law, disappear.

Indeed, taking into account how unit lengths quantum mechanically stretch in a gravitational 
field, the second author obtained the {\it precession of the perihelion of Mercury} as well as the {\it deflection of
light} passing near a celestial body \cite{typl}. Typically, these have been considered to be the
best {\it proofs} of the validity of Einstein's general theory of relativity.
 
 A consequence of our theory is that black holes 
 of macroscopic objects {\it solely due to the force of gravity} do not exist. Rather, when a sufficient amount of
mass coalesces in space, the object becomes either invisible or nearly 
invisible due to the extreme red-shift near such a body.
We thus predict that very dark objects, but no black holes,
 should be found in the center of many galaxies. On the other
hand, if a sufficient amount of mass coalesces causing a total collapse to values of $r$ so small that
other elementary forces become predominant, then it becomes plausible that there
will be a partial or even a total annihilation of the macroscopic body with a corresponding large burst of energy. 
This may explain the presence of the recently discovered ``bigest expanse of nothing'', 
a billion light years wide, which is the space that would normally be occupied by
thousands of galaxies. ``No stars, no galaxies, no anything'' \cite{nothing}.

Although our theory produces results that are practically 
the same as those of the General Theory of Relativity, they are only the same 
up to a third order Taylor expansion. For a further discussion of the issues
involved and how a quantum theory of gravity becomes possible in this setting, see \cite{TY0}, \cite{typl}
and the references therein. 
  Ultimately, the value of any theory rests not upon the conviction or authority of its authors, but on
the fruits of its predictions and its ability to encompass and explain experimental results. 
 
\section*{Acknowledgements} 
 
     The first author thanks Dr. Guillermo Romero, Academic Vice-Rector, and Dr. Reyla Navarro, Chairwomen of
   the Department of Mathematics, at the Universidad de Las Americas for continuing 
   support for this research. He and is a member of SNI 14587.   
   The second author is grateful to Dr. V. Rozanov, Director of the Laser Plasma Theory Division, Lebedev 
   Institute, Russia, to Dr. C. Marchal, Directeur Scientifique de l'ONERA, France, to, Dr. Alwyn Van 
   der Merwe, Editor, Foundations of Physics, to Dr. O. Sinanoglu from Yale University, to Dr. S. Kocak 
   from Anadolu University, and to Dr. E. Hasanov from Isik University; without their sage understanding 
   and encouragement, the seeds of this work could never have born fruit.

    \end{document}